# Noise-Powered Disentangled Representation for Unsupervised Speckle Reduction of Optical Coherence Tomography Images

Yongqiang Huang[1], Wenjun Xia[1], Zexin Lu[1], Yan Liu[1], Hu Chen[1], Jiliu Zhou[1], Leyuan Fang[2], and Yi Zhang[1,*]

[1]College of Computer Science, Sichuan University, Chengdu 610065, China

[2]College of Electrical and Information Engineering, Hunan University, Changsha 410082, China

[*]yzhang@scu.edu.cn

## Abstract:

Due to its noninvasive character, optical coherence tomography (OCT) has become a popular diagnostic method in clinical settings. However, the low-coherence interferometric imaging procedure is inevitably contaminated by heavy speckle noise, which impairs both visual quality and diagnosis of various ocular diseases. Although deep learning has been applied for image denoising and achieved promising results, the lack of well-registered clean and noisy image pairs makes it impractical for supervised learning-based approaches to achieve satisfactory OCT image denoising results. In this paper, we propose an unsupervised OCT image speckle reduction algorithm that does not rely on well-registered image pairs. Specifically, by employing the ideas of disentangled representation and generative adversarial network, the proposed method first disentangles the noisy image into content and noise spaces by corresponding encoders. Then, the generator is used to predict the denoised OCT image with the extracted content features. In addition, the noise patches cropped from the noisy image are utilized to facilitate more accurate disentanglement. Extensive experiments have been conducted, and the results suggest that our proposed method is superior to the classic methods and demonstrates competitive performance to several recently proposed learning-based approaches in both quantitative and qualitative aspects.

## 1. Introduction

Optical coherence tomography (OCT) is employed as a safe and effective diagnostic tool for diverse ocular diseases [1, 2], such as age-related macular degeneration (AMD) and diabetic macular edema (DME), due to its noninvasive imaging character, depth capacity and cross-sectional viewing of tissue structures. However, the speckle noise introduced by the low coherence interferometry imaging process significantly degrades imaging quality, which will seriously affect the subsequent analysis and impedes its clinical application [3]. Therefore, efficient OCT image denoising methods are urgently required.

Over the past decades, a large number of methods have been presented for OCT speckle noise reduction. By improving of the light source, hardware-based approaches reduce the noise of the detector and scanner to some extent, but the speckle or white noise in the imaging system cannot be eliminated [4, 5]. Software-based approaches are the mainstream of OCT image denoising and can be roughly divided into several categories with some overlapping [6]. Reconstruction-based methods [7, 8] process noisy images with handcrafted smoothness priors in the spatial domain. Filtering-based methods usually depend on local or global statistical modeling of the speckle noise within OCT images [9]. Typically, the nonlocal means method (NL-Means) [10] uses a predefined search window to perform nonlocal weighted averaging over noisy images as well as the block-matching and 3D filtering method (BM3D) [11] performs the collaborative filtering over stacked 3D similar patches extracted from the noisy image. The block matching and 4D collaborative filtering (BM4D) method [12] extends BM3D to 3D image volumes. However, these methods need laborious efforts of parameter tuning for different noise levels [13]. Transform-based methods [14, 15] process the degraded OCT images in transform domains, such as frequency or wavelet domain. In spite of impressive denoising results, any unexpected artifacts appear in the transform domain will spread to the whole image. Recently, dictionary learning is utilized in sparsity-based methods for OCT image denoising, such as multiscale sparsity-based tomographic denoising (MSBTD) [16] and nonlocal weighted sparse representation (NWSR) [17]. These methods can suppress noise efficiently, but the denoised images are often oversmoothed, which causes some clinically meaningful details to be lost.

Deep convolutional neural networks (CNNs) demonstrate powerful potential in various vision-related tasks, such as image classification [18], image recognition [19] and image restoration [20]. With a large number of well-registered image pairs (the noisy image and its clean counterpart), supervised learning methods can achieve promising denoising results [20,21]. However, it is difficult to directly transplant these methods to OCT image speckle reduction, since clean OCT images are hard to acquire in clinical practice. Registering and averaging several OCT images acquired at the same position of the same subject is an alternative solution to obtain clean data for supervised learning methods. Based on this operation, different CNN-based methods have been proposed recently [6,22,23], and the results are fairly good. However, due to unconscious body jitter or eye movement during sampling, the obtained OCT images used for averaging might not be captured from the exact same place. As a result, registration is particularly challenging; some motion artifacts may appear, and some critical information may be lost in the averaged result [17,24]. Injecting simulated noise into averaged clean images is a possible solution to collect well-matched image pairs for training [25]. Nevertheless, these methods may not work well in practice since the noise in OCT images does not obey any specific statistical distribution.

To overcome the aforementioned problems, in this paper, we present an unsupervised learning method for OCT image speckle reduction based on Disentangled Representation and Generative Adversarial Network (DRGAN). The proposed model can predict the clean counterparts of the input noisy OCT images without well-registered noisy and clean OCT image pairs in an end-to-end manner. As depicted in Fig. 1, the proposed method takes advantage of the assumption that a noisy OCT image x consists of content and noise components, while the clean image y only has the content part. In this case, we can use a couple of encoders, $E_C$ and $E_N$, to extract the content features $F_C^*$ and noise features $F_N^*$ (* denotes noisy or clean image in corresponding domain), respectively, and then generate the denoised image only with content feature $F_C^*$ by a clean image generator $G_C$. Moreover, the noise patches cropped from the original noisy OCT image are employed to enhance the model capability by performing adversarial learning between the extracted noise and the estimated noise. Once the model is well-trained, we can get the denoised image by feeding the original noisy image into the content encoder $E_C$ and the clean image generator $G_C$ successively.

The main contributions of the paper are summarized as follows: 1) The proposed method is unsupervised that does not rely on the paired OCT images. It eases the strict requirement for averaging and registration and is of great clinical value. 2) Different from traditional methods that assume a specific noise distribution (such as additive Gaussian noise) of OCT images, we extract noise patch without any structure information in the noisy OCT images to facilitate more accurate disentanglement. 3) We experimentally verify our proposed method on two clinical datasets, and the results demonstrate that our approach achieves competitive performance with several state-of-the-art algorithms, including three recently proposed supervised and unsupervised methods.

The rest of the paper is organized as follows: section 2 introduces two kinds of works, unsupervised OCT image denoising algorithms and disentangled representation, which are related to our proposed method. Section 3 elaborates on the proposed method, including the problem formulation, the network architecture and the loss function. The experimental results and corresponding discussion are given in section 4. Section 5 concludes this paper, and future works are suggested.

## 2. Related works

### 2.1 Unsupervised OCT Image Denoising

Since it is impractical to acquire paired images in clinics, the normal way to obtain clean OCT image is to register and average several B-scans that are repeatedly acquired at the same position [6]. However, this image generation strategy has following problems: 1) it will lead to a long acquisition time; 2) unconscious eye movement during scanning makes the registration particularly challenging, and inaccurate registration and averaging may result in motion artifacts and loss of some critical structure information; and 3) some small pathological area could be blurred or totally removed after averaging [26]. Therefore, unsupervised OCT image speckle reduction remains a challenging and urgent task that needs more attention.

Recently, several methods that originate from CycleGAN [27] have been proposed to reduce the speckle noise in OCT images. For example, Manakov et al. proposed HDcycleGAN [28], which modifies the original CycleGAN and treats image denoising as a domain adaption problem between image domains with high and low noise levels. In addition, Guo et al. proposed SNR-GAN [26], which incorporates the structural similarity index measure (SSIM) loss and original CycleGAN, to achieve better performance and structural preservation. The results of those methods are impressive and provide inspiring solutions for unsupervised OCT image denoising.

## 2.2 Disentangled Representations

The basic idea of disentangled representation is to disentangle an image into different domains, which aims at modeling the factors of data variations in an unsupervised manner. Recently, many efforts have been made for learning disentangled representations. Lee et al. [29] proposed to embed images onto domain-invariant content space and domain-specific attribute space to achieve diverse image-to-image translation results. Lu et al. [30] performed single-image deblurring by disentangling the content and blurry features from a blurred image using content encoders and blur encoders, respectively. Sanchez et al. [31] utilized the disentangled representations to learn the shared common information representation and the exclusive specific information representation of satellite image series. Liao et al. [32] introduced disentangled representations to the metal artifacts reduction of computed tomography images, and the results look satisfactory.

## 3. Proposed method

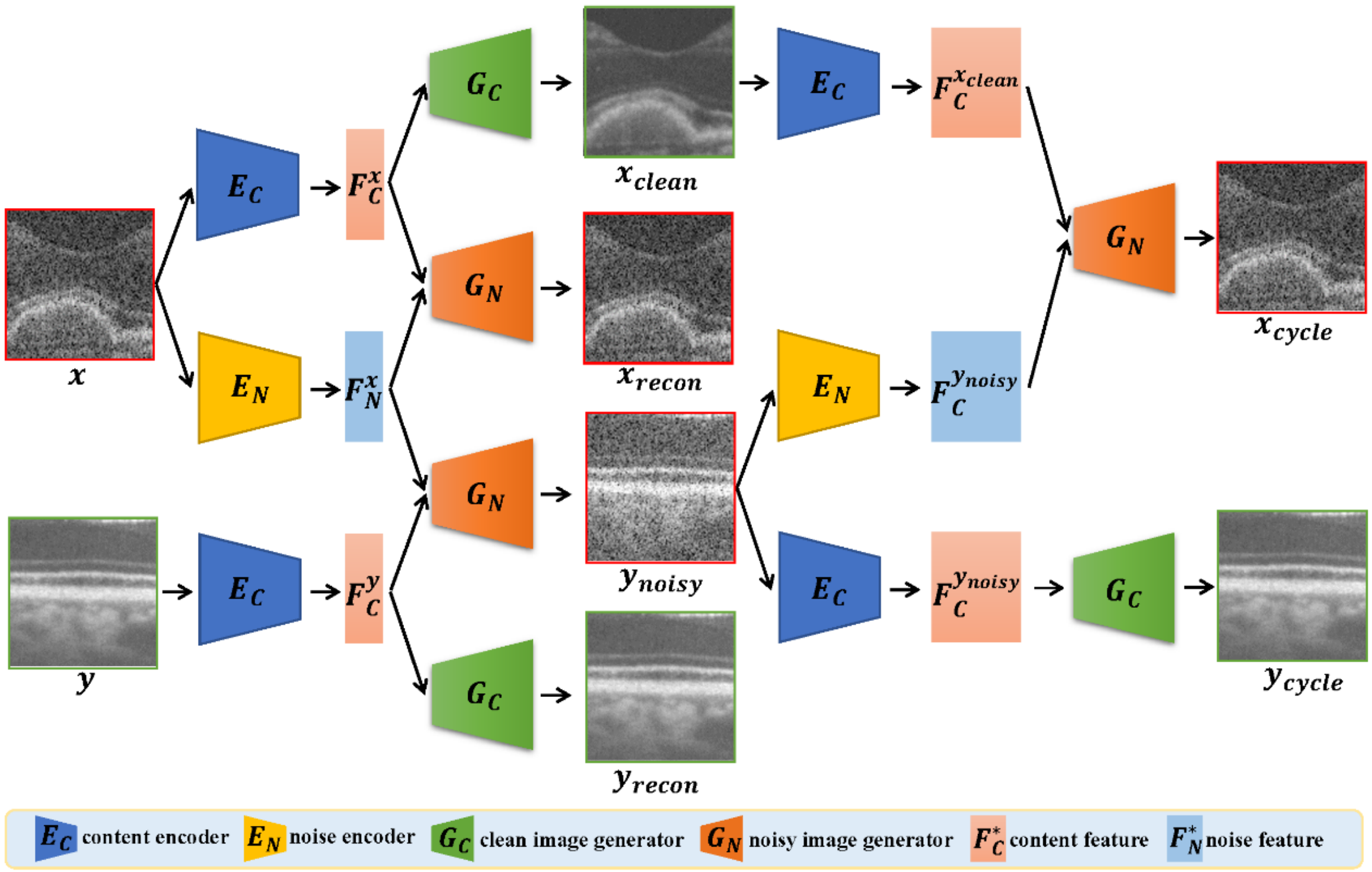

Fig. 1. Framework of our proposed method.

### 3.1OCT Speckle Reduction Model

Given two unpaired images: a noisy image $x \in I^n$ and a clean image $y \in I^c$, the goal of our proposed DRGAN is to learn a denoising model from the noisy image domain $I^n$ to the clean image domain $I^c$ and reconstruct the clean observation of $x$. To achieve this purpose, we employ the idea of disentangled representation that assumes that the images in $I^n$ can be disentangled into a domain-specific content space and a domain-invariant noise space and those in $I^c$ only can be transformed to the content space. Fig. 1 demonstrates the framework of our proposed DRGAN. It contains a series of encoders and generators: $E_C$, $E_N$, $G_C$ and $G_N$. The encoders attempt to decompose an image sample from the image domain to the corresponding content or noise space, while the generators map the encoded features in the content or/and noise space back to the image domain.

Specifically, images from different domains are first encoded as the domain-specific content features $F_C^*$ and domain-invariant noise features $F_N^*$ by the content encoder $E_C$ and the noise encoder $E_N$, respectively:

$$F_C^x = E_C(x), F_N^x = E_N(x), F_C^y = E_C(y)$$

Then, the clean image generator $G_C$ is utilized to reconstruct a clean image from a content feature $F_C^*$, which means that decoding from $F_C^x$ will remove the speckle noise from $x$, and decoding from $F_C^y$ will reconstruct $y$.

$$x_{clean} = G_C(F_C^x), y_{recon} = G_C(F_C^y)$$

In addition, the noisy image generator $\mathrm{G_N}$ is used to generate noisy observations, which takes $F_C^y$ and $F_N^x$ as inputs and is expected to generate the noisy observation of $y$. Similarly, while feeding $F_C^x$ and $F_N^x$, $\mathrm{G_N}$ will reconstruct $x$. They can be formulated as,

$$y_{noisy} = G_N(F_C^y, F_N^x), x_{recon} = G_N(F_C^x, F_N^x)$$

Moreover, in order to handle unpaired data, the disentanglement is also performed on the generated images $x_{clean}$ and $y_{noisy}$, and decoding from the corresponding features, we can obtain the cycle noisy image $x_{cycle}$ and the cycle clean image $y_{cycle}$ by $G_N$ and $G_C$, respectively:

$$F_C^{x_{clean}} = E_C(x_{clean}), F_C^{y_{noisy}} = E_C(y_{noisy}), F_N^{y_{noisy}} = E_N(y_{noisy})$$

$$x_{cycle} = G_N(F_C^{x_{clean}}, F_N^{y_{noisy}}), y_{cycle} = G_C(F_C^{y_{noisy}})$$

Once the disentanglement is well-addressed and the model is well-trained, we can get the speckle-reduced OCT images simply by propagating the noisy images forward to $E_C$ and $G_C$ sequentially.

## 3.2Training losses

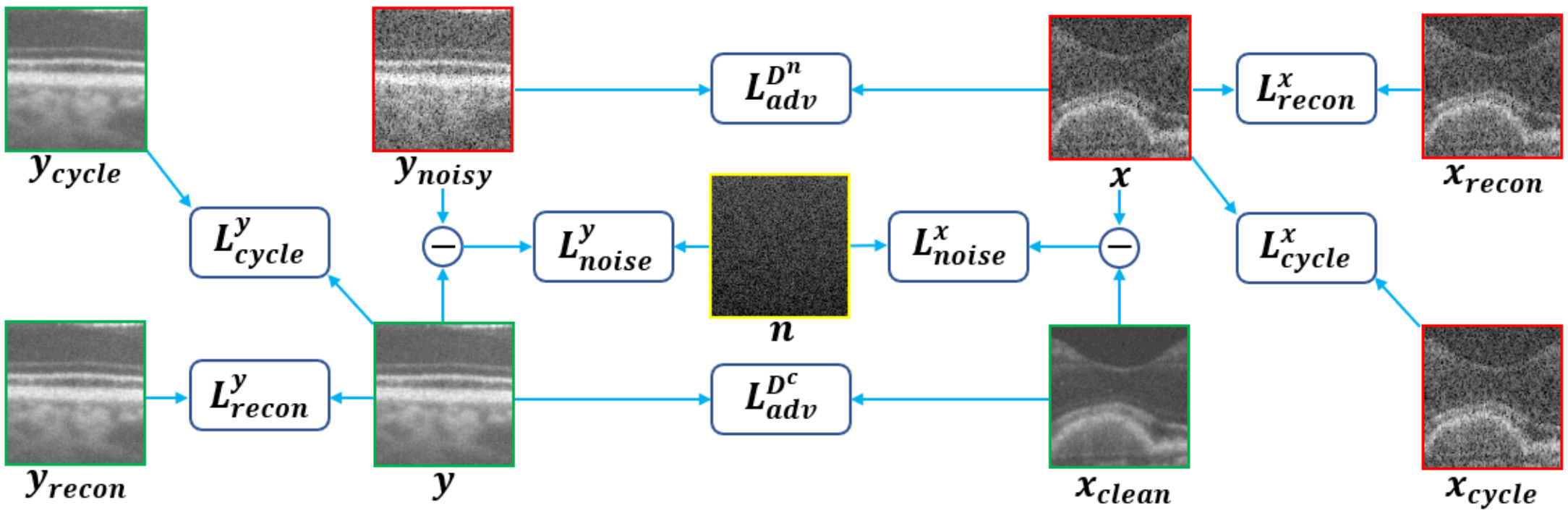

Fig. 2 The main components of the loss function.

Fig. 2 illustrates the main components of our proposed loss function and their relationships with the inputs and outputs of the network. To obtain an accurate disentanglement of the content and noise features from a noisy input, four parts are included as our loss function, namely, the domain adversarial loss $L_{adv}$, the reconstruction loss $L_{recon}$, the cycle-consistency loss $L_{cycle}$ and the proposed novel noise loss $L_{noise}$, which employs the noise patches for adversarial learning. The overall loss function of our proposed DRGAN is as follows:

$$L = L_{adv} + \lambda_1 L_{cycle} + \lambda_2 L_{recon} + \lambda_3 L_{noise}$$

where $\lambda_1$, $\lambda_2$ and $\lambda_3$ are the weighting coefficients that control the impacts of each component, and these components are elaborated in the following subsections.

### 3.2.1 Domain Adversarial Loss

For the sake of generating the clean counterpart of $x$ or the noisy counterpart of $y$, two discriminators, noisy domain discriminator $D_N$ and clean image domain discriminator $D_C$, are employed to ensure the generated image belongs to similar distribution as the image in the corresponding domain in an adversarial way. Specifically, $D_C$ takes $y$ and the generated image $x_{clean}$ as inputs to determine which one is the real clean image and which is generated by $G_C$. Likewise, $D_N$ is utilized to distinguish real noisy image $x$ and the fake noisy image $y_{noisy}$ generated by $G_N$. Therefore, the domain adversarial loss is defined as follows:

$$L_{adv}^{\mathrm{I}^c} = \mathbb{E}[logD_C(y)] + \mathbb{E}[1 - logD_C(x_{clean}))]$$

$$L_{adv}^{\mathrm{I}^n} = \mathbb{E}[logD_N(x)] + \mathbb{E}\big[1 - logD_N(y_{noisy}))\big]$$

$$L_{adv} = \arg\min_{E,G}\max_{D}\,(L_{adv}^{\mathrm{I}^c} + L_{adv}^{\mathrm{I}^n}),$$

where $\mathbb{E}[\cdot]$ denotes the expectation operator.

### 3.2.2 Cycle-Consistence Loss

Inspired by CycleGAN [27], a cycle-consistency loss is employed to ensure that the generated image $x_{cycle}$ owns the same content as $x$, and similarly, $y_{cycle}$ possesses the same content as $y$. The cycle-consistency loss is defined as follows:

$$L_{cycle} = \mathbb{E}\left[\left\|x - x_{cycle}\right\|_1\right] + \mathbb{E}\left[\left\|y - y_{cycle}\right\|_1\right]$$

where $||\cdot||_1$ represents the $l_1$-norm, and the cycle loss enforces the constraint that $x_{cycle} \approx x$ and $y_{cycle} \approx y$.

### 3.2.3 Reconstruction Loss

To generate the clean counterpart of $x$ or the noisy counterpart of $y$, $G_C$ and $G_N$ are expected to reconstruct the input images, $x$ and $y$, respectively. To achieve this, we apply the following reconstruction loss to facilitate $x_{recon} \approx x$ and $y_{recon} \approx y$:

$$L_{recon} = \mathbb{E}[||x - x_{recon}||_1] + \mathbb{E}[||y - y_{recon}||_1]$$

### 3.2.4 Noise Loss

Since the noise in OCT images has more than one source, it contains multiple types of noise, such as the scanning and electronic noise of the imaging device, and the speckle noise of the interferometric imaging modality, which does not obey a specific statistical distribution. Based on this consideration, it is unreasonable to restrict the noise distribution in OCT images using the common measurements (i.e., KL divergence). To conquer this obstacle, we propose to leverage the noise patch extracted from background part of the noisy image to constrain that the estimated noise removed from $x$ or injected to $y$ shares similar statistical distribution as the noise patch. Specifically, a noise discriminator $D_{PN}$ is employed to distinguish the noise patch from the estimated noise $x - x_{clean}$ or $y_{noisy} - y$ as follows:

$$L_{adv}^{x} = \mathbb{E}[logD_{PN}(n)] + \mathbb{E}[1 - logD_{PN}(x - x_{clean})]$$

$$L_{adv}^{y} = \mathbb{E}[logD_{PN}(n)] + \mathbb{E}[1 - logD_{PN}(y_{noisy} - y)]$$

$$L_{adv} = \arg\min_{E,G}\max_{D}(L_{adv}^{x} + L_{adv}^{y}),$$

### 3.3 Network Architectures

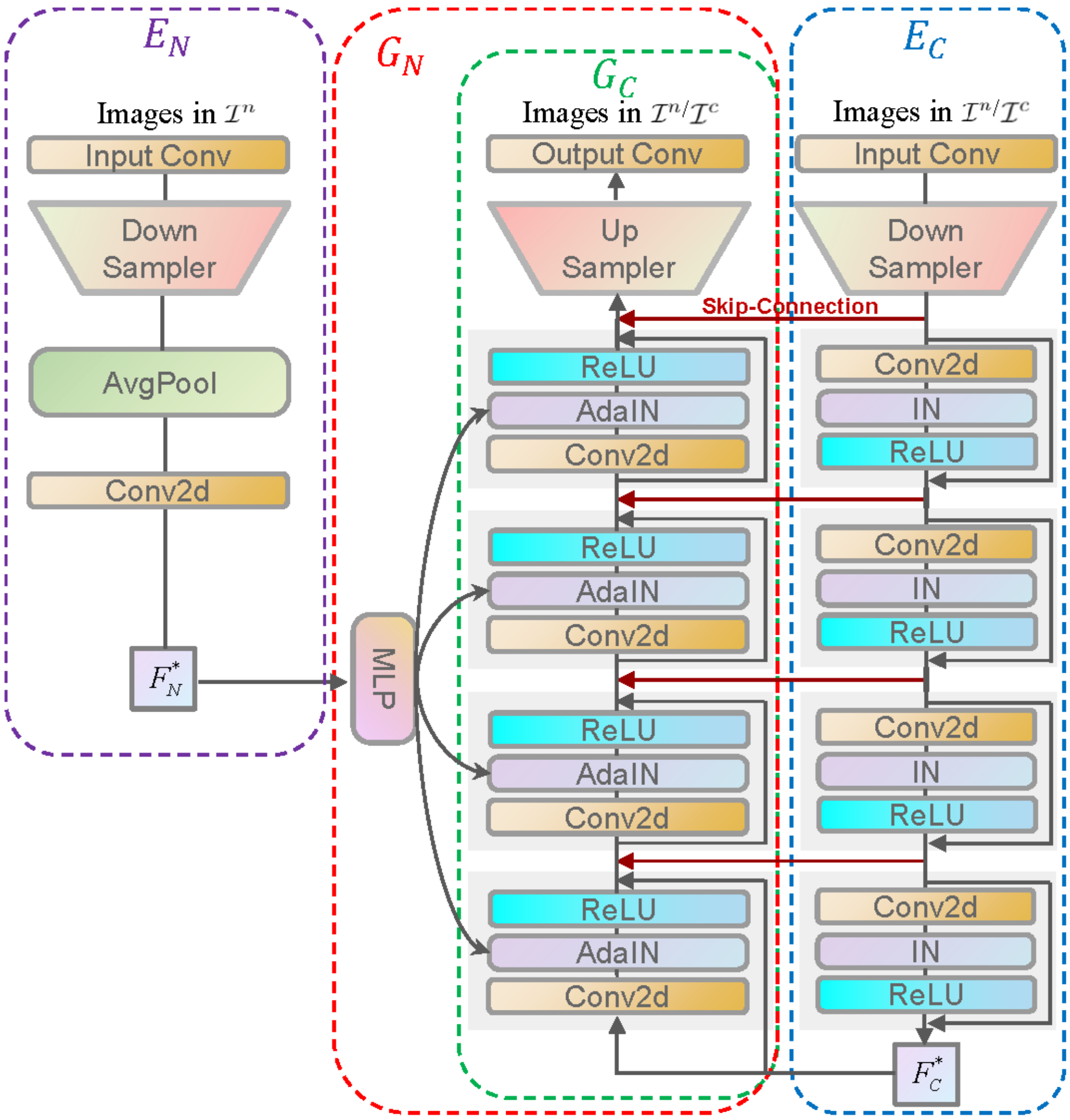

Fig. 3 Network architecture of the encoders and generators of DRGAN

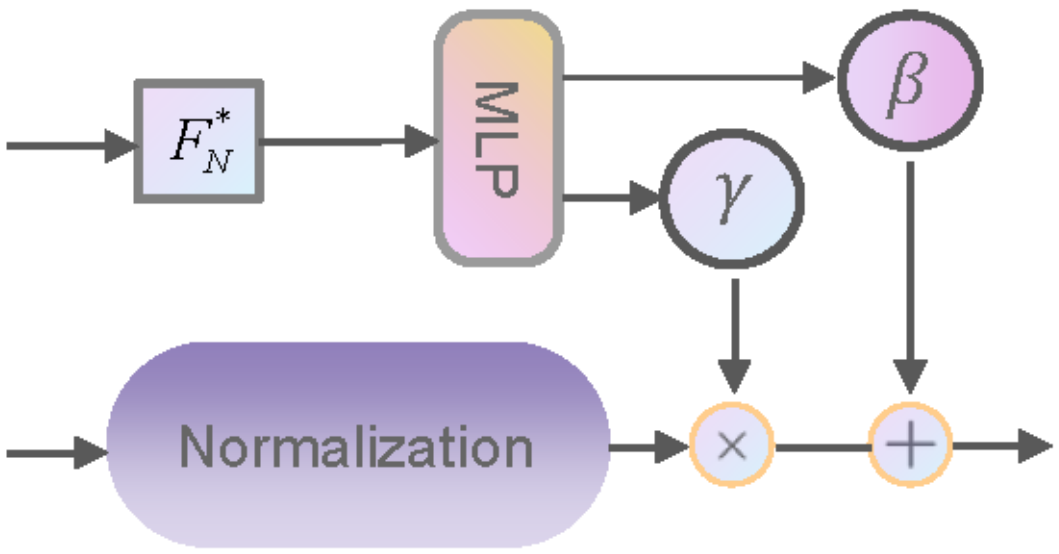

Fig. 4 AdaIN block

The architecture of the encoders and generators in DRGAN are illustrated in Fig. 3, from which we can observe that content encoder $E_C$ consists of an input convolutional layer, a down sampler and four residual blocks while the noise encoder $E_N$ has one input convolutional layer, a down sampler and an adaptive average pooling layer followed by a 1x1 convolutional layer. The generators are not simple decoders as oppose to the encoders' architectures, and their structures vary for generating clean and noisy image respectively. For the clean image generator $G_C$, the skip-connections between corresponding layers of $E_C$ and $G_C$ are utilized to fuse the low- and high-level features, which are more conductive for the preservation of semantic details and makes the network converge quickly [33]. In terms of noisy image generator $G_N$, we borrow the ideas of SPADE [34] and AdaIN [35] and propose to inject the noise feature into the adaptive instance normalization layer in each residual block by Multilayer Perceptron (MLP), which has been proven effective to maintain more semantic information in the generated image[34]. Specifically, as depicted in Fig. 4, the extracted noise feature is fed into MLP to learn the affine parameters $\gamma$ and $\beta$ that are used to normalize the previous output feature maps as follows:

$$\gamma = \sigma\left(MLP(F_N^*)\right)$$

$$\beta = \mu\left(MLP(F_N^*)\right)$$

$$AdaIN(f_i, F_N^*) = \gamma \frac{f_i - \mu(f_i)}{\sigma(f_i)} + \beta$$

where $f_i$ is the feature map of the i-th channel, and $\mu(f_i)$ and $\sigma(f_i)$ are the mean and variance of $f_i$, respectively. Specifically, $f_i$ is first normalized to the distribution with mean 0 and variance 1, and then the noise is injected by multiplying $\gamma$ and adding $\beta$, which is similar to the style generator in [35]. The process of nonlinear mapping and affine transform from the latent noise space by MLP can be regarded as sampling the noise from the learned distribution $F_N^*$. For the discriminator, we simply employ the same architecture of that in PatchGAN[36].

# 4. Experimental design and results

## 4.1 Data preparation

We combined two clinical datasets to validate our proposed model. The first one was originally introduced in [16], and we employed 17 eye SDOCT image pairs from 17 subjects with and without AMD. In addition, our experiments also included 28 450×900 (height × width) eye image pairs from 28 patients in the second dataset [37].

It is easy to notice that the original noisy OCT images have large areas only containing speckle noise, and no retinal or layer structure information exists (see the region just below the white line in Fig. 5(a)). Therefore, without any automatic segmentation algorithm, as shown in Fig. 5(a), we can roughly divide a noisy sample into two parts: the information part (above the white line) and the background part (below the white line). Then, we used a sliding window with size of 256 × 256 to traverse the background part and obtained noise patches indicated by the yellow rectangles in Fig. 5(a) for the noise loss in $L_{noise}$.

As mentioned above, the clean OCT images in both datasets were acquired by registering and averaging multiple samples obtained at the same position of the same subject. This operation may lead to oversmoothed or artifact-affected results, which will cause unnecessary problems for OCT image denoising methods. This will be discussed in detail in section 4.5.1. As a result, we manually removed 19 pairs of OCT images of which the clean counterpart was oversmoothed or the image details were destroyed during the registering and averaging procedures. Fig. 5 (b) shows a case of this problem, especially in the regions located in the blue box. After that, 26 images pairs remained, and we random selected 10 pairs as the training set and the remaining 16 pairs as the test set. More specifically, all the images were center-cropped to 450 × 900 (height × width). For setting the unsupervised training, the 10 training images were divided into two parts; the first five noisy images were regarded as the images in $\mathrm{I}^{n}$, and the last five clean images were the images in $\mathrm{I}^{c}$. These unpaired five noisy images and five clean images were traversed with a window with size of 256 × 256 and a stride of 8 to obtain total 9680 unpaired noisy patches (the red rectangles in Fig. 5(a)) and clean patches (the green rectangles in Fig. 5(a)).

## 4.2 Implementation

Our model was implemented with PyTorch, and all the experiments were conducted on a Ubuntu 18.04 operation system and a NVIDIA GTX 1080Ti GPU. In the training stage, the Adam optimizer was adopted with $\beta_1 = 0.5$ and $\beta_2 = 0.999$, and the learning rate was set to 1e-4 for all 100 epochs. Additionally, we experimentally set the weighting hyperparameters $\lambda_1 = 10$, $\lambda_2 = 10$ and $\lambda_3 = 1$ in the loss functions for all cases.

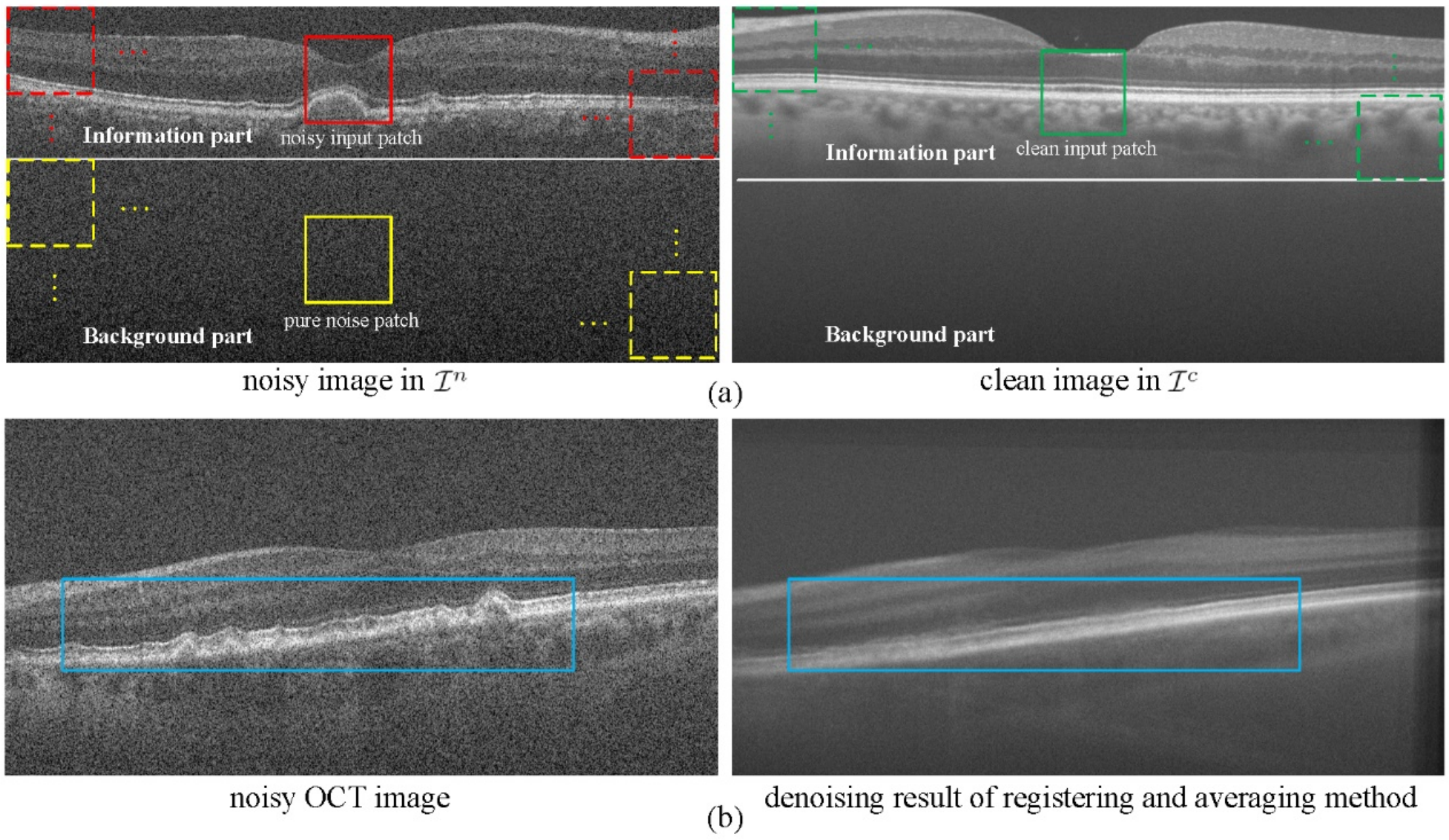

Fig. 5 Data preparation. (a) An illustration of information/background part division and the acquisition process of the noisy, clean and noise patches. (b) A case where the registering and averaging operation fails to denoise but leads to an oversmoothed result on the retinal layers, and these kind of samples are not included in our dataset.

## 4.3 Baselines and Quantitative Measurements

To evaluate the performance of the proposed model quantitatively and qualitatively, several state-of-the-art algorithms, including Median [38], NL-Means [39], Bilateral [40], Wavelet [14], BM3D [11], SNR-GAN [26], NWSR [17], and Edge-sensitive cGAN [6], were included in comparison. The first six methods can be roughly treated as unsupervised methods, and the last two are supervised. For the first five classical methods, we used the build-in implementation in the Python Scikit-image package to verify the denoising performance on 16 test data. The last three learning-based approaches were implemented according to the original papers, and they were trained and tested on the same data as utilized in the proposed DRGAN. Four unsupervised metrics, including contrast-to-noise ratio (CNR), edge preservation index (EPI), mean-to-standard-deviation ratio (MSR), and equivalent number of looks (ENL), which respectively evaluate the image contrast, the ability of detail preservation, the denoising performance and the smoothness of the background part, were employed as quantitative metrics. Following the common practice [17,37], we manually selected several regions of interest (ROIs) to calculate the metrics. Since the boundaries of the retinal layers are viewed as the most important part of OCT images to sense disease severity and pathogenic processes [41], four regions (the red rectangles #1~4 in Figs. 6 and 7) were selected at or near the retinal layers to denote the signal ROIs, and one background ROI is selected at the homogeneous region (the green rectangle #0). Thus, the quantitative metrics can be defined as follows:

$$CNR = \frac{1}{m}\sum_{i=1}^{m}\left[10\log_{10}\left(\frac{\mu_i - \mu_b}{\sqrt{\sigma_i^2 + \sigma_b^2}}\right)\right]$$

$$EPI = \frac{\sum_i \sum_j |I_d(i+1,j) - I_d(i,j)|}{\sum_i \sum_j |I_n(i+1,j) - I_n(i,j)|}$$

$$MSR = \frac{1}{m}\sum_{i=1}^{m}\frac{\mu_i}{\sigma_i}$$

$$ENL = \frac{{\mu_b}^2}{\sigma_b^2}$$

where $\mu_b$ and $\sigma_b$ are the mean and standard deviation of the selected background ROI, respectively, $\mu_i$ and $\sigma_i$ denote the mean and standard deviation of the $i\text{-}th$ signal ROI, respectively, and $m$ stands for the number of signal ROIs. $I_d$ and $I_n$ in EPI represent the information part of the denoised and noisy images respectively, and *i* and *j* denote the *i-th* row and *j-th* column of the image, respectively.

## 4.4 Results

To validate the visual effects of the proposed method, two typical images were selected in Fig. 6, in which we compared our method with unsupervised methods, including six classic algorithms and one unsupervised learning-based method. Four signal ROI regions are chosen and magnified for better visual inspection. It is easy to notice that the results of the proposed DRGAN method are obviously better than all the other methods in both speckle reduction and edge preservation. Specifically, all the compared method can suppress the noise to different levels. It is noticeable that heavy noise still remains in the results of Median filtering and Bilateral filtering, and Median filtering blurs the edge details. NL-Means and Wavelet introduced extra unexpected artifacts, which are evident in the enlarged ROIs. BM3D and SNR-GAN method achieved better performance in speckle reduction, but they still tended to destroy the edge structure near the retinal layers as shown in the ROIs. Our method eliminated most of the noise, and the structures in retinal layers are quite clear.

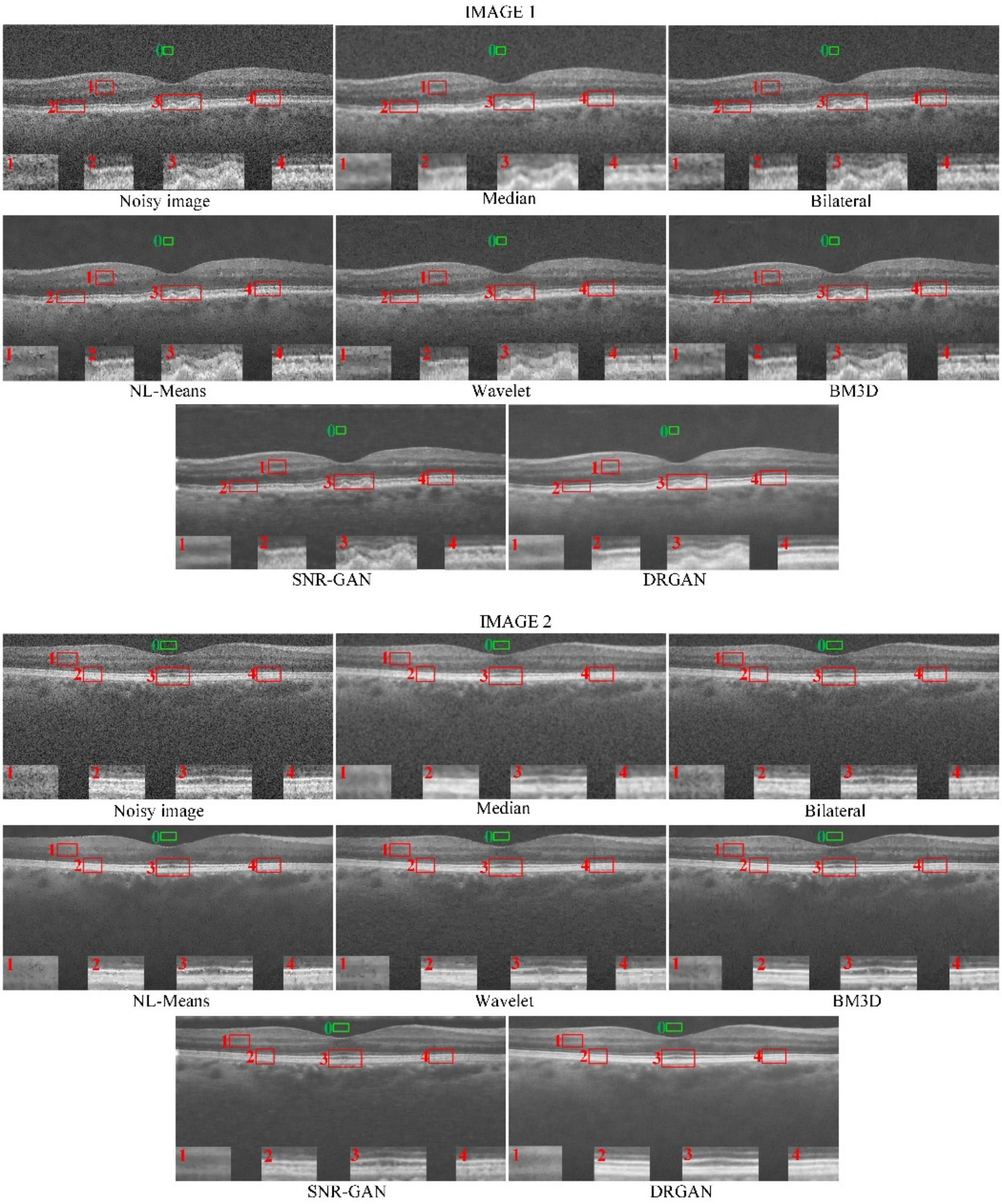

Fig. 6. Visual comparison of our proposed method with other unsupervised methods.

Fig. 7 visually compares our proposed DRGAN method with two state-of-the-art supervised learning-based methods. It can be observed that both methods achieved satisfactory speckle reduction performance in the homogeneous region, but edge-sensitive cGAN led to a blurring effect on the image details, as shown in the selected ROIs. These results occur likely because the generator of edge-sensitive cGAN was originally designed to process an image whose size is equal to an integer power of 2, such as 256, 512, and so on, and an extra padding operation is needed when applying it on our test image with size of 450×900. NWSR achieved better performance than edge-sensitive cGAN and obtained sharper edges in retinal layers. However, some layers were lost and some structures were destroyed when comparing our results and the registered and averaged results, as depicted in ROIs #2-4 in Fig. 7.

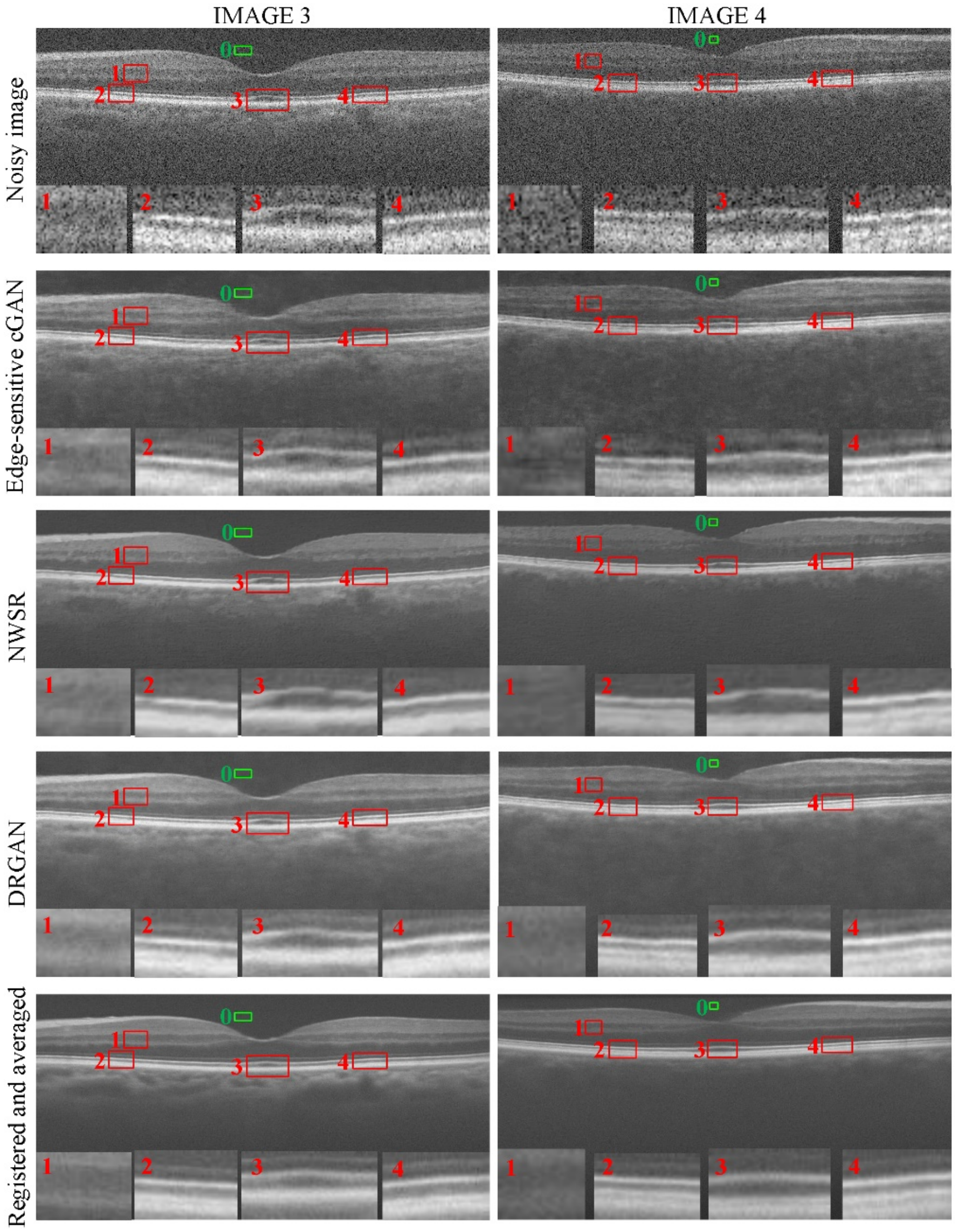

Fig. 7. Visual comparison of our proposed method with two state-of-the-art supervised learning-based methods.

Table 2 demonstrates the quantitative results of different approaches, obtained by calculating the mean values of four metrics on all 16 test images. It can be seen that our proposed DRGAN method achieves the best results in terms of CNR and EPI among all the methods. Our method shows great potential for contrast and edge preservation, which are of great importance in clinical diagnosis. The MSR value of our method ranks second among all the methods and is very close to the best score achieved by NWSR, which indicates that the proposed DRGAN method can effectively suppress the speckle noise in retinal layer regions. It is also noticed that the value of ENL of our method only lies in the middle of all the methods. Since ENL evaluates the smoothness in the homogeneous region, it can be interpreted that BM3D, NWSR and SNR-GAN achieved more smoothed results in the background part than DRGAN. However, it is a common sense that the greatest challenge in image denoising is to balance the tradeoff between noise suppression and detail preservation rather than eliminate all the noise. This is consistent with the visual results in Figs. 6 and 7 in which BM3D, NWSR and SNR-GAN produced oversmoothed results to different degrees.

Table 2. Quantitative results of all the methods

| | | CNR | MSR | EPI | ENL |
|---|---|---|---|---|---|
| unsupervised | Median | 2.9756 | 4.6907 | 0.8174 | 135.2064 |
| | NL-Means | 2.6380 | 4.3204 | 0.8948 | 303.5238 |
| | Bilateral | 2.6020 | 4.1466 | 0.9412 | 61.1037 |
| | Wavelet | 2.9034 | 4.4624 | 0.3912 | 175.5959 |
| | BM3D | 2.7165 | 4.2996 | 0.9715 | **637.2720** |
| | SNR-GAN | 2.6051 | 4.4118 | 0.8377 | 390.7385 |
| | Ours | **3.1877** | 4.7464 | **0.9862** | 317.4043 |
| supervised | NWSR | 2.9969 | **4.8602** | 0.8921 | 584.8610 |
| | Edge-sensitive cGAN | 2.6368 | 4.3516 | 0.9434 | 156.1844 |

## 4.5 Robustness Analysis

### 4.5.1 Registering and averaging denoising

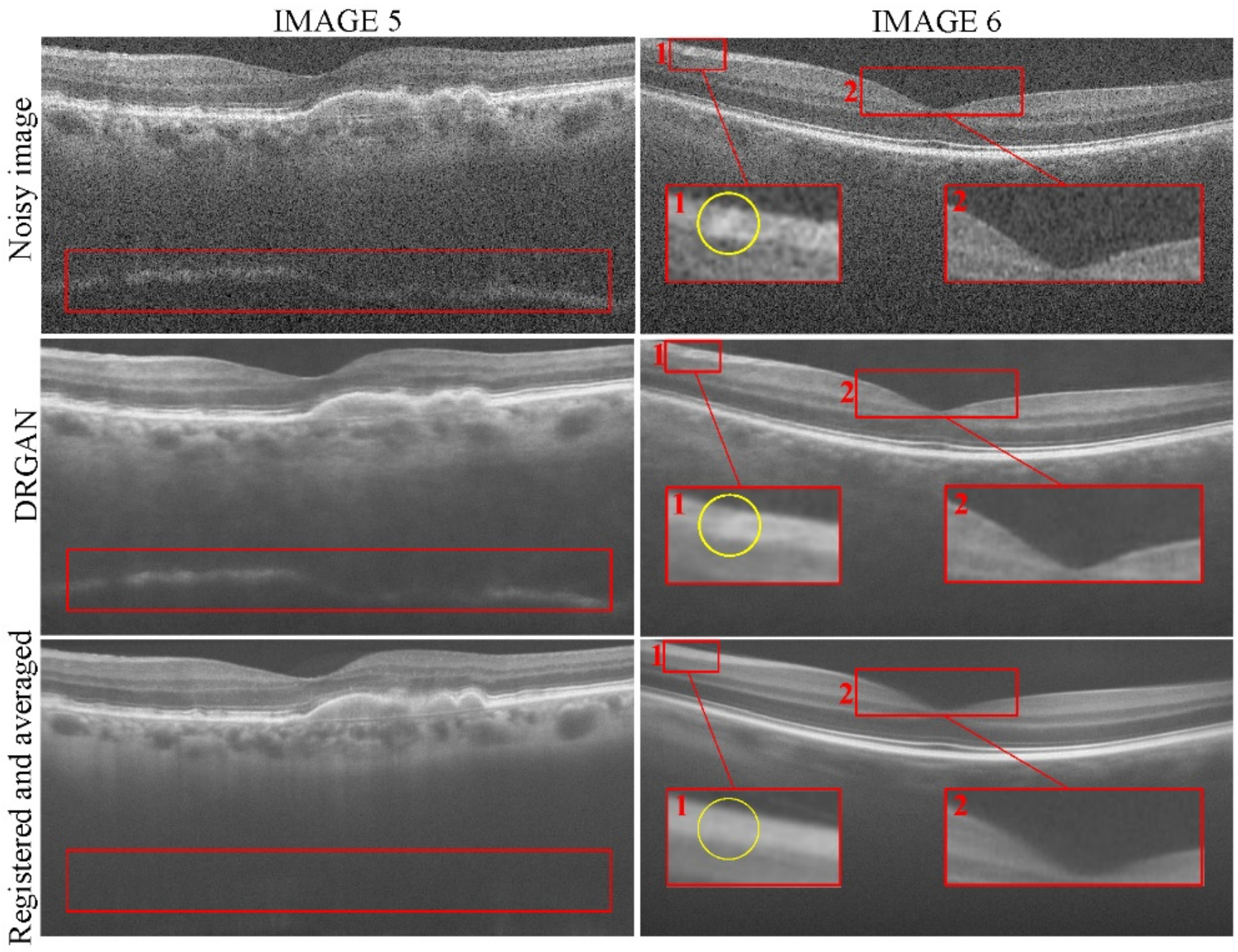

Fig. 8 Visual comparison of the proposed method and the registering and averaging denoising approach.

As aforementioned, the most common way to obtain a noiseless OCT image in commercial scanners is to acquire a volume of OCT images at the same position of the same subject and average these images to suppress the speckle noise within the noisy OCT images. This process seems easy to implement, but it takes strict conditions to get good results. Specifically, the averaging process requires sample multiple OCT images at the same position. Unfortunately, due to unconscious body jitter or eye movement during scanning, OCT images are not always captured from the exact same place, which challenges the registering and averaging operations to a great extent. If the results are not accurate enough, motion artifacts may be introduced and some details will be lost [17,24]. As a result, in practice, a low-sampling rate is often adopted to accelerate the data acquisition process and reduce the influence of unconscious motion blur. However, the spatial resolution of reconstructed images obtained with a low sampling rate will degrade and have a negative impact on the clinical value for diagnosis. Fig. 8 demonstrates two typical cases processed by the proposed method and the common registering and averaging approach. We can easily notice that in the red rectangle in IMAGE 5, the denoising method embedded in the commercial scanner loses deep layer information, while our method preserves these structures well. The same phenomenon is also occurred in the region indicated by a yellow circle in IMAGE 6. A possible lesion is erased after registering and averaging operations. In addition, these operations may also result in oversmoothed results, as depicted in the regions indicated by red rectangles in IMAGE 6.

#### 4.5.2 Effectiveness of noise loss

Since noise loss is a major contribution of our proposed method, an ablation study is performed to verify its effectiveness. Fig. 9 shows the visual results of one typical case processed by our proposed DRGAN model with/without noise loss, and the mean values of four quantitative metrics calculated on all 16 test images are listed in Table 3. From Fig. 9, the results processed by DRGAN without noise loss show that the noise still remains, and the edges are blurred, which are also proved by the quantitative result as shown in Table 3. In contrast, the result obtained by DRGAN with noise loss not only achieve better speckle reduction performance but also maintain the edges well.

#### 4.5.3 Noise distribution assumption

To further verify the effectiveness of our proposed noise loss, we also conducted the experiments assuming that the speckle noise obeys Gaussian distribution. Under this assumption, we followed the common practice [29,30,42] using KL divergence loss to regularize the distribution of noise feature $F_N^*$, which is encoded by $E_N$ and is drawn from the Gaussian distribution as $F_N^* \sim N(0,1)$. Fig. 9 shows that by imposing the constraint that $F_N^*$ obeys the Gaussian distribution, the model removes most noise and maintains more details; however, there are still speckles remaining, which can be sensed in ROI #2 and ROI #4. The quantitative results listed in Table 3 support our observations.

Table 3 Quantitative results of different noise assumptions.

| | CNR | EPI | MSR | ENL |
|---|---|---|---|---|
| DRGAN w noise loss | 3.1877 | 0.9862 | 4.7464 | 317.4043 |
| DRGAN w/o noise loss | 1.9450 | 0.9432 | 4.1914 | 197.6196 |
| Gaussian noise assumption | 2.9856 | 0.9786 | 4.4942 | 207.2773 |

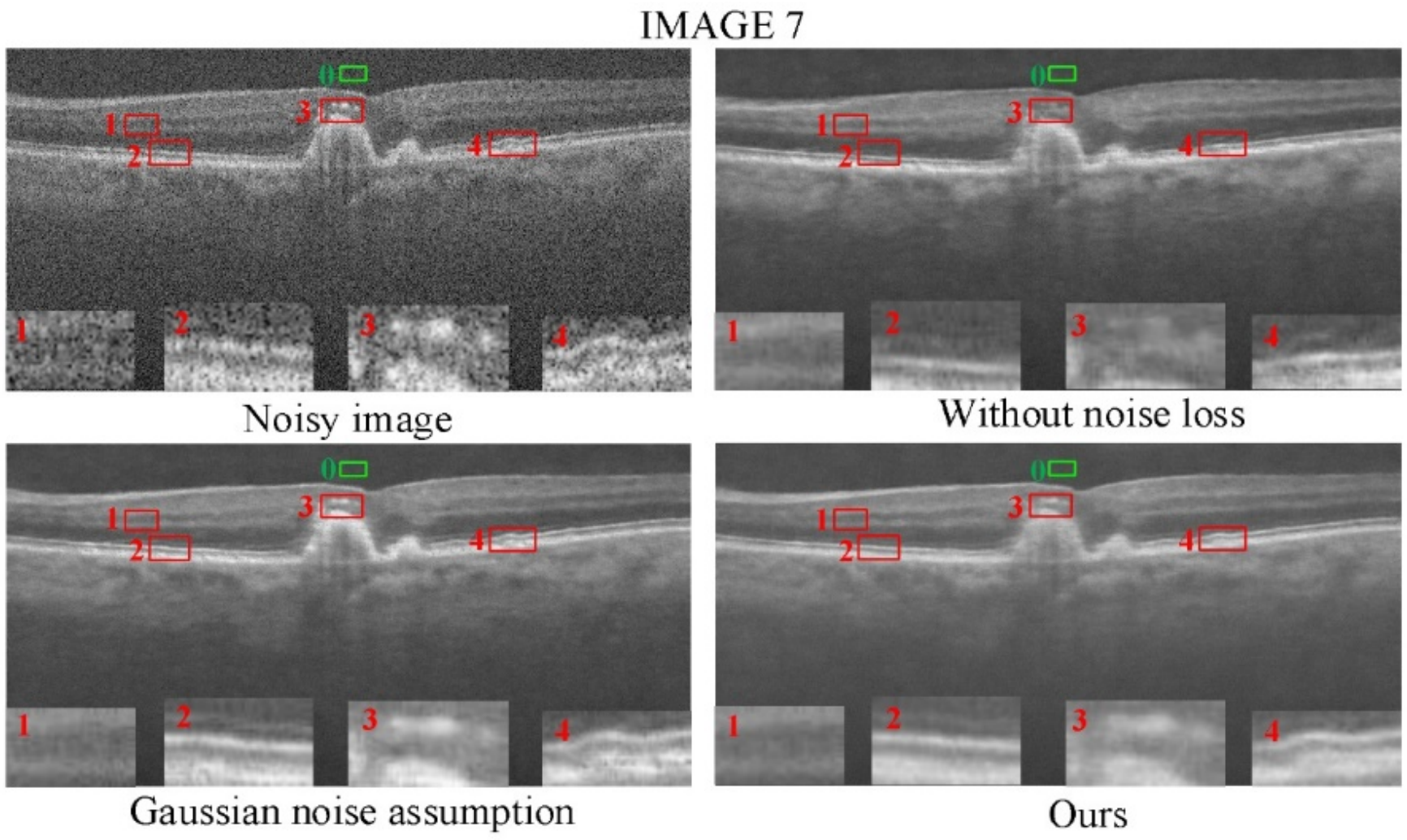

Fig. 9. Visual comparison of different noise assumptions.

### 4.5.4 Effectiveness for Layer Segmentation

In retinal OCT images, the segmentation of layers that contain various anatomical and pathological structures is crucial for the diagnosis and study of ocular diseases [42]. As a result, the denoised images are expected to not only retain the clinical important structures, but also make the segmentation results more accurate. To verify this, 100 OCT images in [43] were first denoised by all the methods and then a publicly available OCT segmentation and evaluation GUI (OCTSEG) was utilized to segment the layers in the denoised images. Fig. 10 shows the results of a typical case, from which we can see that the segmentation performance of all the denoised images are improved compared to the result of original noisy image. Among all the methods, our result achieves the best performance of the automatic layer segmentation, since the segmentation lines are not crossed and the boundaries between layers are obvious, especially in the selected two regions.

## Conclusion

In this paper, we propose a novel unsupervised OCT image speckle reduction method integrating the ideas of adversarial learning and disentangled representation. Utilization of noise and content disentanglement of an OCT image using a corresponding encoder and generator allow for impressive results. Additionally, we qualitatively and quantitatively compared our method with several state-of-the-art OCT denoising approaches, including five classic methods and three learning-based methods. The results indicate that our proposed DRGAN method favorably outperforms other methods in noise reduction and detail preservation.

In our future work, we will try to further optimize the proposed SRGAN for clinical applications by evaluating the denoising performance on the task driven datasets from different scanners and scanning protocols. Meanwhile, an adaptive parameter selection strategy for the weights of different loss components will be helpful.

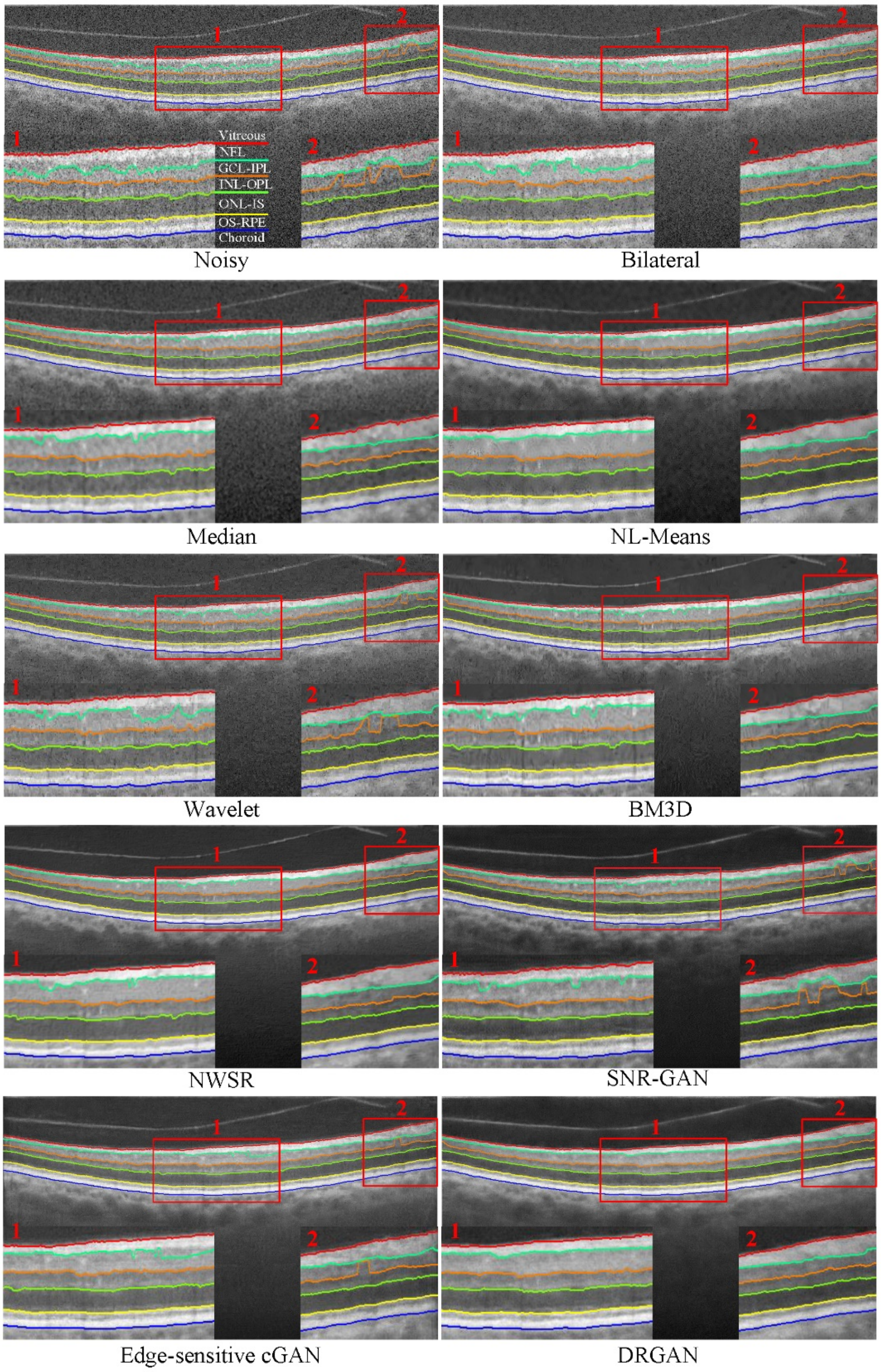

Fig. 10. Visual comparison of layer segmentation performance on the denoised images processed by all the methods.